# Demonstration of Si-doped Al-rich thin regrown Al(Ga)N films on AlN on sapphire templates with >$10^{15}$/cm$^3$ free carrier concentration using close-coupled showerhead MOCVD reactor

Swarnav Mukhopadhyay, Parthasarathy Seshadri, Mobinul Haque, Shuwen Xie, Ruixin Bai, Surjava Sanyal, Guangying Wang, Chirag Gupta, Shubhra S. Pasayat

Department of Electrical and Computer Engineering, University of Wisconsin-Madison, Madison, WI, 53706, USA

## Abstract:

Thin Si-doped Al-rich (Al>0.85) regrown Al(Ga)N layers were deposited on AlN on Sapphire template using metal-organic chemical vapor deposition (MOCVD) techniques. The optimization of the deposition conditions such as temperature, V/III ratio, deposition rate, and Si concentration resulted in a high charge carrier concentration (> $10^{15}$ cm$^{-3}$) in the Si-doped Al-rich Al(Ga)N films. A pulsed deposition condition was employed to achieve a controllable Al composition greater than 95% and to prevent unintended Ga incorporation in the AlGaN material deposited using the close-coupled showerhead reactor. Also, the effect of unintentional Si incorporation on free charge carrier concentration at the regrowth interface was studied by varying the thickness of the regrown Al(Ga)N layer. A maximum charge carrier concentration of 4.8×$10^{16}$ /cm$^3$ and 7.5×$10^{15}$ /cm$^3$ were achieved for Al$_{0.97}$Ga$_{0.03}$N and AlN films with thickness <300 nm compared to previously reported n-Al(Ga)N films with thickness ≥400 nm deposited using MOCVD technique.

High-performance power devices with voltage handling capabilities of several kVs are currently in high demand in the power electronics industry. The ultrawide bandgap (UWBG) semiconductors such as Al-rich Al(Ga)N have shown numerous promise for high-voltage operation[1–6]. Similarly, Al(Ga)N-based deep ultra-violet (DUV) emitters, such as light emitting diodes (LEDs) and laser diodes (LDs) are also showing promising performance[7–10]. However, the performance enhancement of both the power and optoelectronic devices requires adequate control of doping into the Al-rich AlGaN films. The Al-rich AlGaN films with Al content >85% suffer from the low conductivity of the film due to higher activation energy of n and p-type dopants, such as Si and Mg, respectively[11–15]. Furthermore, the Si-doped Al-rich Al$_x$Ga$_{1-x}$N (x>0.85) films tend to show self-compensation and DX center formation of the dopants which limits the availability of the free charge carrier in the film[16–18]. It is also reported that the cation vacancy ($V_{III}$) complex formation with Si ($V_{III}$-Si), which behaves like an acceptor trap state is also a conductivity limiting factor[11,16]. Apart from that, the Al(Ga)N deposited using the metal-organic chemical vapor deposition (MOCVD) technique suffers from the presence of carbon (C), which is known to act as a compensating acceptor ($C_N$)[11]. It was demonstrated that the optimized deposition condition for reducing the formation energy of the $V_{III}$-Si complex requires a lower deposition temperature, whereas the reduction of the $C_N$ formation in the Al(Ga)N needs a high deposition temperature[19]. So, a trade-off must be created to obtain highly conductive films. Si-doped AlN films (Al-polar and N-polar) deposited using molecular beam epitaxy (MBE) exhibit high free carrier concentration (>$10^{18}$/cm$^3$) due to their metal-rich growth condition and low deposition temperature compared to MOCVD, reducing the $V_{III}$-Si formation and enhancing the free-carrier concentration[12,20]. However, using a similar deposition method with MOCVD is challenging.

Although a significant amount of research has been done on the deposition technique of Si-doped Al-rich (>85%) Al(Ga)N material using vertical and horizontal MOCVD reactors, very few reports have been published for the deposition technique of Al-rich Al(Ga)N materials using close-coupled showerhead (CCS) reactors[21,22]. The CCS reactors are widely accepted for commercially producing III-nitride films and are easily scalable[23]. However, the CCS reactors are known to have issues with unintentional Ga incorporation[23,24], which makes it difficult to obtain a controllable Al composition >95% in AlGaN. Since AlN has low n-type conductivity, depositing 95-97% AlGaN may offer better n-type conductivity while minimizing lattice mismatch with AlN buffer layers or the substrate. Also, AlGaN thin films with Al composition between 10-95% have very low thermal conductivity compared to AlN or GaN due to the dominant phonon-alloy scattering phenomena[25,26]. The phonon-alloy scattering decreases exponentially as the Al composition becomes >95%, thus the thermal conductivity of AlGaN enhances rapidly[26]. Using AlGaN films (Al > 95%) with higher thermal conductivity improves heat dissipation, benefiting power electronic devices and deep UV emitters.

In both power electronics and DUV emitters, low-resistive Al-rich n-AlGaN is required to obtain ohmic contacts[27,28]. It has been observed that obtaining ohmic contact with the Al-rich (>0.85) AlGaN is non-trivial[29]. So, to obtain low specific contact resistivity (<1 µΩ.cm$^2$), it is desired to have a significantly high n-type dopant concentration (>$10^{19}$/cm$^3$) such as Si in the Al-rich n-AlGaN layer. However, it has been observed that with an increase in Si doping concentration, the material quality can degrade and introduce v-pits and dislocations[30], which is not ideal for both power electronic and optoelectronic devices. So, developing a low-resistance Al-rich n$^{++}$-AlGaN layer with smooth surface morphology is crucial for improving the efficiency of the AlGaN-based devices.

In previous studies, regrown III-nitride materials using MOCVD have shown high unintentional Si incorporation at the regrowth interface[31,32]. This high unintentional Si concentration (>$5\times10^{19}$ /cm$^3$) can behave as deep acceptor levels causing reduced free carrier concentration in Si-doped AlN[33,34]. So, reducing the unintentional Si incorporation or its effect would help to increase the free charge carrier concentration in regrown Si-doped thin Al-rich Al(Ga)N films.

This study demonstrated an optimized deposition condition of regrown n-type Si-doped Al-rich Al(Ga)N films on an AlN/sapphire template using a CCS MOCVD reactor to maximize free electron concentration. The effect of deposition conditions on the free electron concentration was determined. A controllable Al composition beyond 95% was observed to be difficult to achieve in the CCS reactor due to unintentional Ga incorporation. Even after increasing the Al-containing precursor flow while keeping the Ga-containing precursor flow constant, the Al composition did not increase. Thus, for obtaining 97% AlGaN, a pulsed deposition condition was pursued, rather than a continuous deposition condition. Furthermore, an unintentional Si incorporation was observed at the interface of the AlN template and the regrown n-Al(Ga)N films, which affected the electrical performance of the n-Al(Ga)N films. A thicker Al(Ga)N layer (>300 nm) exhibited a higher free charge carrier concentration compared to a thinner layer (<100 nm). The total thickness of the Al(Ga)N layer was kept below 300 nm as obtaining a higher conductivity film with a lower thickness would be cost-effective. Finally, this work showed state-of-the-art free carrier concentration with the thinnest layer of regrown n-AlN on AlN/Sapphire template compared with the reported results from n-AlN deposited on different substrates such as bulk AlN, SiC and Sapphire.

Al-rich Al(Ga)N films (65±5 nm) were deposited on AlN/sapphire templates in the CCS MOCVD reactor using triethylgallium (TEGa), trimethylaluminum (TMAl) as group-III precursors and ammonia ($NH_3$) as group-V precursor. Silane was used as a gaseous precursor for Si doping. First, different deposition conditions such as deposition temperatures (1050 °C, 1100 °C, 1150 °C, and 1210 °C), V/III ratio (530, 750 and 3000), deposition rate (0.52 Å/s, 0.7 Å/s, and 1 Å/s), and Si concentration ($4\times10^{19}$ /cm$^3$, $6\times10^{19}$ /cm$^3$, and $8\times10^{19}$ /cm$^3$) were used to find the optimized deposition condition to maximize free carrier concentration of n-AlN (Table 1). The V/III ratio was varied by increasing the $NH_3$ flow while keeping the TMAl flow constant. To adjust the deposition rate of AlN, the TMAl flow rate was altered while maintaining a constant V/III ratio. Next, the TMAl/TEGa ratio was varied to obtain different Al compositions of $Al_xGa_{1-x}N$ ($0.57 \leq x \leq 1$) (Table 2). An Al composition over 95% was achieved with pulsed deposition, continuously supplying TMAl and $NH_3$, while pulsing TEGa with a 4-second ON/OFF cycle. $H_2$ served as the carrier gas for Al(Ga)N deposition. Finally, a thickness series of $Al_{0.97}Ga_{0.03}N$ and AlN with thicknesses up to 280 nm was performed to understand the effect of the unintentional Si incorporation at the interface on free charge carrier concentration.

Table 1. Optimization of maximum charge carrier concentration in Si-doped AlN (65±5 nm)

| Experiments | Temperature (°C) | V/III ratio | Deposition Rate (Å/s) | Si (/cm$^3$) | $n_s$ (/cm$^3$) |
|---|---|---|---|---|---|
| $E_{A1}$ | 1050 | 750 | 0.7 | $6\times10^{19}$ | $4\times10^{13}$ |
|  | 1100 |  |  |  | $1\times10^{14}$ |
|  | 1150 |  |  |  | $1.5\times10^{14}$ |
|  | 1210 |  |  |  | N/A |
| $E_{A2}$ | 1150 | 530 | 0.7 | $6\times10^{19}$ | $6\times10^{13}$ |
|  |  | 750 |  |  | $1.5\times10^{14}$ |
|  |  | 3000 |  |  | $7\times10^{13}$ |
| $E_{A3}$ | 1150 | 750 | 0.52 | $6\times10^{19}$ | $2\times10^{13}$ |
|  |  |  | 0.7 |  | $1.5\times10^{14}$ |
|  |  |  | 1 |  | $1.51\times10^{14}$ |
| $E_{A4}$ | 1150 | 750 | 0.7 | $4\times10^{19}$ | $1\times10^{14}$ |
|  |  |  |  | $6\times10^{19}$ | $1.5\times10^{14}$ |
|  |  |  |  | $8\times10^{19}$ | N/A |

Table 2. Variation of TMAl/TEGa ratio for obtaining different Al composition

| Experiments | Temperature (°C) | V/III ratio | Deposition Rate (Å/s) | Si (/cm$^3$) | TMA/TEGa ratio | Al% |
|---|---|---|---|---|---|---|
| $E_{B1}$ | 1150 | 750 | 0.7 | $6\times10^{19}$ | $\infty$ | 100 |
| $E_{B2}$ | | | ~1 | | 12.69 | 97 (Pulsed) |
| $E_{B3}$ | | | ~1 | | 12.69 | 95 |
| $E_{B4}$ | | | 0.92 | | 6.43 | 93 |
| $E_{B5}$ | | | 0.88 | | 2.87 | 86 |
| $E_{B6}$ | | | 1.12 | | 0.64 | 57 |

The composition of the Al(Ga)N was measured using omega-2theta and reciprocal space mapping (RSM) techniques with the high-resolution X-ray diffraction (XRD) Panalytical Empyrean tool. The surface roughness of the Al(Ga)N films was measured by atomic force microscopy (AFM) using Bruker Icon AFM in tapping mode. The carrier concentration of the n-Al(Ga)N films was measured using mercury CV and Hall measurements using Lake Shore MCS-EMP-4T-V and the Toho HL9980 hall measurement systems. The Si concentration and Al/Ga ratio were measured by the secondary ion mass spectroscopy (SIMS) method using the Cameca SIMS measurement tool.

The Si-doped AlN films deposited at different temperatures showed a prominent trend of how the trade-off between $V_{III}$ and $C_N$ can be obtained in Table 1 ($E_{A1}$). An increase in the temperature from 1050 °C to 1150 °C, increased the free charge carrier concentration from $4\times10^{13}$/cm$^3$ to $1.5\times10^{14}$/cm$^3$. We speculate that this 3.75 times increase in the charge concentration may have been related to the reduction of the $C_N$, which acts as an acceptor-type trap. At lower deposition temperatures, C incorporation becomes very high in MOCVD-deposited III-nitride materials due to the formation of C–group-III bonds[35–37]. The presence of the methyl groups in the TMAl was the main reason for the C incorporation. Increasing the temperature promotes the elimination of C from the group-III element by breaking the C–group-III bonds. A sharp decrease in the carrier concentration was observed when the deposition temperature was increased beyond 1150 °C. It was demonstrated previously in the literature that with increasing deposition temperature, the $V_{III}$ formation increases[22,36,37]. So, the probability of $V_{III}$-Si complex formation increases rapidly. Thus, the sharp decrease in carrier concentration was presumably related to the $V_{III}$-Si complex formation, like the trend observed by *Washiyama et al.*[36,37]. As the maximum charge was observed at 1150 °C, the rest of the experiments were performed at 1150 °C.

In the next set of experiments, the V/III ratio was varied from 530 to 3000 to identify the effect of $C_N$ and N vacancy formation on the free charge carrier concentration, as shown in Table 1 ($E_{A2}$). At a low V/III ratio (530), the N vacancy formation probably increased which enhanced the C incorporation in the N sites, forming $C_N$[36]. It caused a lower charge carrier concentration of $6\times10^{13}$/cm$^3$. When the V/III ratio was increased from 530 to 750, the charge increased by 2.5 times, reaching $1.5\times10^{14}$/cm$^3$, indicating less C$_N$ formation. However, a high V/III ratio (3000) decreased the charge carrier concentration again from $1.5\times10^{14}$/cm$^3$ to $7\times10^{13}$/cm$^3$. We speculate that the reduction in the charge carrier concentration was related to the increased group-III vacancy formation with increased NH$_3$[36].

Next, the effect of the deposition rate on the charge carrier concentration was observed, shown in Table 1 (E$_{A3}$). The increase in deposition rate from 0.57 Å/s to 0.72 Å/s showed an increase in the charge concentration from $2\times10^{13}$/cm$^3$ to $1.5\times10^{14}$/cm$^3$. We speculate that an increment in deposition rate helped to reduce the desorption of group-III species, thus the V$_{III}$ formation got suppressed, and a similar phenomenon has been observed for high-composition n-AlGaN layers as reported by another group in ref. [38]. However, with increasing TMAl flow rate the unintentional C concentration also increases, thus the chances of C$_N$ formation also increase[35,38,39]. So, a very high TMAl flow rate would cause detrimental effects in terms of film conductivity. A further increase in the deposition rate from 0.72 Å/s to 1 Å/s, did not increase the charge carrier concentration significantly, which might be due to the trade-off between reduced V$_{III}$ and increased C$_N$ concentration with increasing TMAl flow rate[38]. Further increments in the TMAl flow rate can increase C$_N$ formation. Thus, the deposition rate was maintained at 0.7 Å/s to avoid excess C$_N$ formation.

Next, a silicon concentration variation was performed to identify the optimal Si concentration needed to obtain the highest charge in the AlN layers as shown in Table 1 (E$_{A4}$). As expected, increasing the Si concentration from $4\times10^{19}$/cm$^3$ to $6\times10^{19}$/cm$^3$ increased the charge concentration from $10^{14}$ /cm$^3$ to $1.5\times10^{14}$ /cm$^3$. However, increasing Si concentration from $6\times10^{19}$/cm$^3$ to $8\times10^{19}$/cm$^3$ decreased the charge concentration sharply, indicating the self-compensation phenomena of Si[36,37]. At very high Si concentrations, the V$_{III}$-Si complex formation, which acts as a compensator, becomes more energetically favorable[36]. Thus, the concentration of the V$_{III}$-Si complex increases at very high Si concentration, causing enhanced compensation of free charge carriers leading to the self-compensation phenomenon of Si[36]. So, a Si concentration of $6\times10^{19}$/cm$^3$ was chosen to be optimal.

After optimizing the deposition condition for the Si-doped AlN, different Al composition AlGaN films were deposited to identify the charge incorporation efficiency in different alloy compositions. It was observed that the charge incorporation efficiency exponentially increased with the increase of Ga mole fraction in AlGaN[40]. Nearly 5 orders of magnitude increase in charge carrier concentration was observed when the Ga mole fraction increased from 0 to 43% (Figure 1). Also, the mobility of the Al(Ga)N film increased from ~7 cm$^2$/Vs to 45 cm$^2$/Vs with increasing Ga mole fraction from 0% to 43%. Obtaining AlGaN with <5% Ga was not trivial in the CCS reactor, as it was observed that the Ga mole fraction saturated at 5%. With increasing TMAl flow by 38%, the Ga mole fraction could not be lowered further while both the TMAl and TEGa precursors were continuously flowing into the deposition chamber (continuous deposition mode). This saturation of Ga incorporation was mostly related to the unintentional Ga incorporation in the reactor chamber, as also reported elsewhere[23,24]. So, for obtaining Al$_{0.97}$Ga$_{0.03}$N, the pulsed deposition condition was used as described in the experimental methods section.

AFM scans showed that with a reduction of Ga composition, surface roughness increased for Al(Ga)N films (65±5 nm) deposited with continuous deposition mode. However, a clear difference in the surface roughness was observed between the continuous deposition mode and the pulsed deposition mode thin films. The pulsed deposition mode showed almost 4 times lower surface roughness (0.42 nm) compared to all the continuous mode samples (trend line) (Figure 2). We speculate the pulsed mode deposition increased the diffusion length of the group-III adatoms which helped to improve the surface roughness[41].

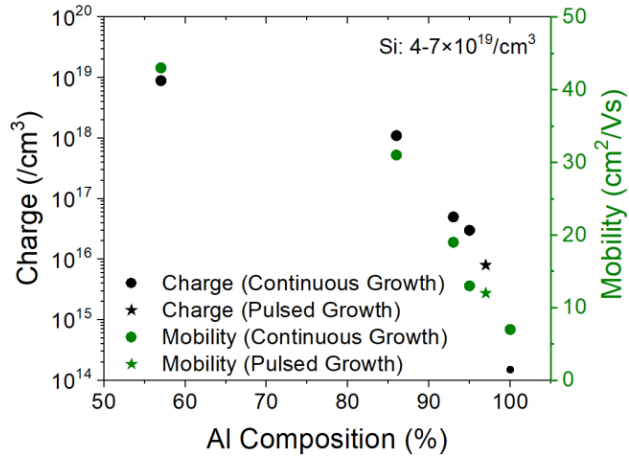

Figure 1. Charge concentration in different Al compositions of Al(Ga)N films (65±5 nm) with a Si concentration between $4-7\times10^{19}$/cm$^3$

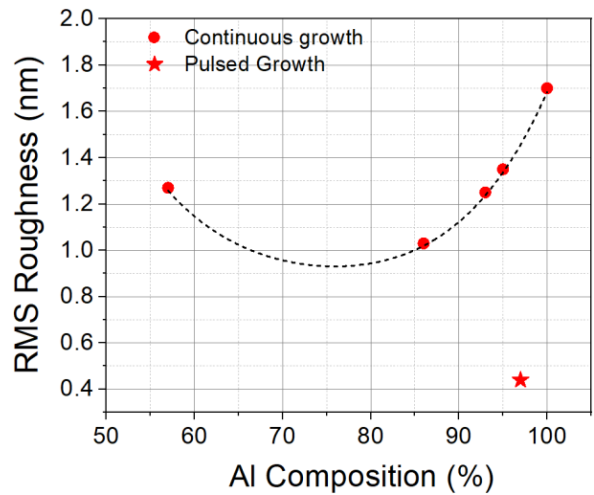

Figure 2. Effect of the surface roughness (AFM scan of 10×10 µm$^2$) of the Al(Ga)N films (65±5 nm) with different compositions of the Al and different deposition modes

The SIMS measurement of Si-doped AlN showed very high unintentional Si incorporation at the interface between the regrown AlN and the AlN template (Figure 3a). This closely resembled the high unintentional silicon incorporation observed in regrown GaN films using MOCVD, as documented in the literature[31,32]. The unintentional Si incorporation was related to the presence of the residual Si in the MOCVD chamber as reported previously[32]. It was hypothesized that this high unintentional silicon incorporation led to carrier compensation[33,34], and therefore in our case, as the thickness of the Al(Ga)N layer increased, the concentration of free charge carriers also increased with lower carrier compensation away from the regrowth interface. This indicated that the Si-doping which was used in our regrown Al(Ga)N layers was optimal and the actual doping efficiency was higher than the measured values for the samples with a lower regrown layer thickness (65±5 nm). The charge concentration increased by ~6 times for $Al_{0.97}Ga_{0.03}N$, with 4 times increase in thickness reaching $4.8 \times 10^{16}/cm^3$, reaching the highest reported free carrier concentration in 97% n-AlGaN. Similarly, for AlN, the charge was enhanced by almost 50 times achieving $7.5 \times 10^{15}/cm^3$ while the thickness was increased by 3 times (Figure 3b). This demonstrates that the compensation effect of the unintentional Si incorporation was more drastic in AlN compared to $Al_{0.97}Ga_{0.03}N$. We think that this phenomenon was related to the higher probability of forming the $DX^{-1}$ state of the Si and $V_{III}$-Si complex in n-AlN, which severely affects the free electron concentration in n-AlN films compared to n-AlGaN[42].

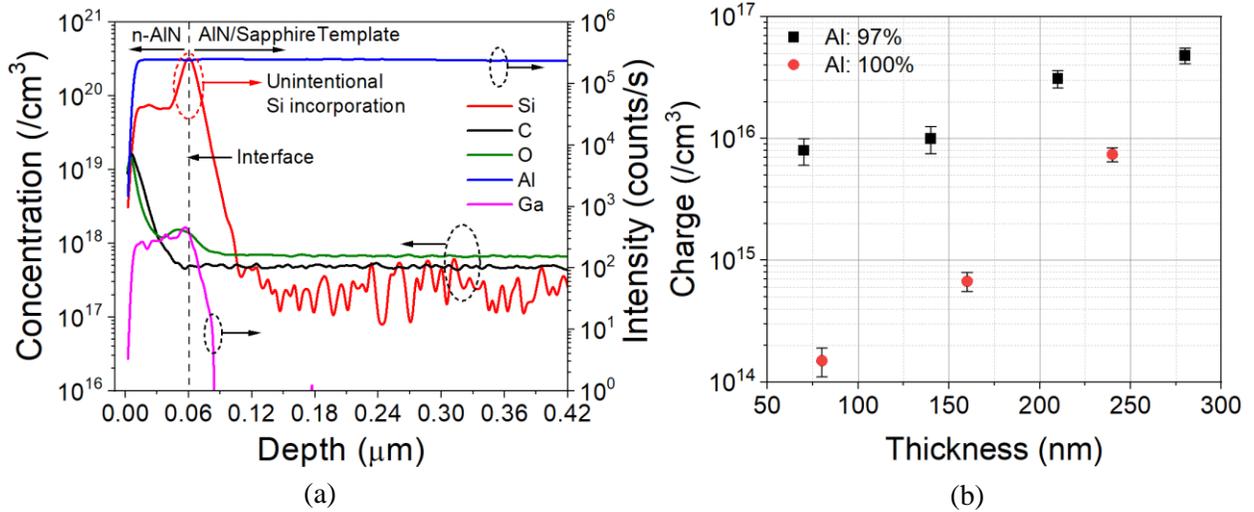

Figure 3. (a) SIMS measurement of Si-doped AlN (65±5nm) film (b) Thickness vs charge concentration for $Al_{0.97}Ga_{0.03}N$ and AlN films

The SIMS measurement of the Al(Ga)N films (65±5 nm) showed that for AlN the Al/Ga ratio was ~1000. This proves that the unintentional Ga incorporation in AlN was (~0.1%), which was below the alloy level [43]. Finally, the AFM measurement of 280 nm n-$Al_{0.97}Ga_{0.03}N$ and 240 nm n-AlN showed smooth surface morphology with a surface roughness of 0.56 nm and 0.13 nm (2 μm×2 μm), respectively (Figure 4). A larger v-pit depth of ~25 nm was observed in n-$Al_{0.97}Ga_{0.03}N$ compared to ~1 nm in n-AlN, which was the reason behind the increased roughness in $Al_{0.97}Ga_{0.03}N$. The large difference in v-pit depth between n-$Al_{0.97}Ga_{0.03}N$ and n-AlN films with similar Si doping concentrations ($6 \times 10^{19}/cm^3$) is still under investigation and might be related to the difference in film thickness.

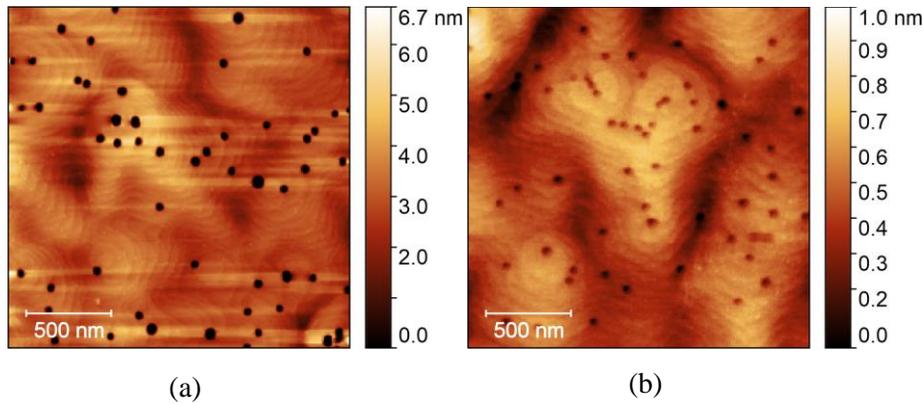

(a)            (b)

Figure 4. AFM image (2 μm×2 μm) of (a) 280 nm n-$Al_{0.97}Ga_{0.03}N$ ($R_a$=0.56 nm) and (b) 240 nm n-AlN ($R_a$=0.13 nm)

The obtained free charge carrier concentration in Si-doped AlN measured using Hall measurement was compared with the state-of-the-art n-AlN films deposited using MOCVD as reported in the literature (Figure 5). A high free carrier concentration of $7.5\times10^{15}$ /$cm^3$ with the lowest AlN thickness (240 nm) was demonstrated in this study (Figure 5). This is especially useful for regrown $n^{++}$ AlN contact development for AlN or Al-rich AlGaN-based power devices. Also, it provides insight into further development of the interface quality at the regrown AlN layer and the AlN on sapphire template junction, to help further increase the free carrier concentration. However, the mobility was still below 10 $cm^2$/V.s, which needs to be improved significantly by reducing the dislocation density, as reported previously[37,44].

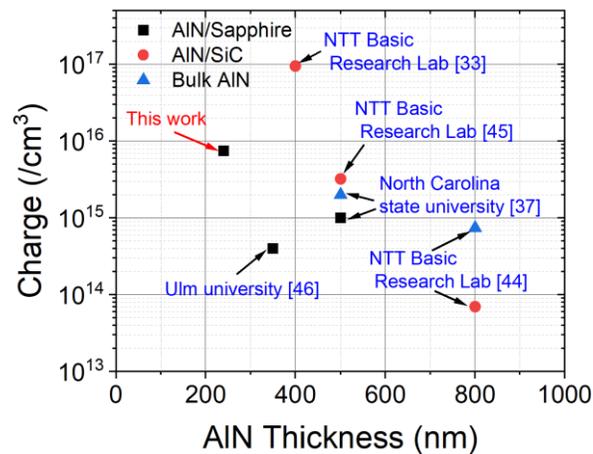

Figure 5. Comparative study of the free charge carrier concentration with AlN thickness for different AlN templates

In conclusion, this study demonstrates methods to obtain high free carrier concentrations in Si-doped high composition (Al>90%) AlGaN using a commercially available CCS reactor by optimizing deposition temperature, V/III ratio, deposition rate, and Si concentration. The pulsed deposition method was used to obtain 97% AlGaN in a CCS reactor with a very low surface roughness of 0.42 nm (scan area of 10×10 μm$^2$) and a charge concentration of 4.8×10$^{16}$ /cm$^3$, which is the highest carrier concentration reported so far. Moreover, the effect of the unintentional Si incorporation at the regrown AlN and AlN on sapphire template interface was explored, and it was found that increasing the thickness of the regrown AlN thickness helps to increase free carrier concentrations. A maximum free carrier concentration of 7.5×10$^{15}$ /cm$^3$ was obtained with 240 nm regrown AlN on the AlN on sapphire templates. Further reduction of DX state formation, Si self-compensation and V$_{III}$-Si complex formation would greatly improve the charge concentration in n-AlN films in the future, which would be crucial for AlN-based power devices and deep UV emitters.

Acknowledgment:

The authors gratefully acknowledge Lake Shore Cryotronics and Toho Technology for providing the Hall measurement results. The research is partially funded by the NSF CAREER (ECCS-2338683), NSF ASCENT (ECCS-2328137) and ARPA-E ULTRAFAST (AWD00001917).